\documentclass[reprint,
superscriptaddress,
showpacs,
amsmath,amssymb,
aps,
prb,
]{revtex4-1}

\usepackage{mathrsfs}
\usepackage{latexsym}
\usepackage{graphicx}
\usepackage{dcolumn}
\usepackage{bm}

\begin{document}

\preprint{APS/123-QED}

\title{Superfluid (Amplitude) Fluctuations Above $T_c$ in a Unitary Fermi Gas}

\author{Hua Li}
\affiliation{Physics Department, Boston College,Chestnut Hill, Massachusetts 02467, USA}
\author{Jason Jackiewicz}
\affiliation{Department of Astronomy, New Mexico State University}
\author{Kevin S. Bedell}
\affiliation{Physics Department, Boston College,Chestnut Hill, Massachusetts 02467, USA}

\date{\today}

\begin{abstract}
 We study the transport properties of a Fermi gas with strong attractive interactions close to the unitary limit. In particular, we compute the spin diffusion lifetime of the Fermi gas due to superfluid fluctuations above the BCS transition temperature $T_c$. To calculate the spin diffusion lifetime we need the scattering amplitudes. The scattering amplitudes are dominated by the superfluid fluctuations at temperatures just above $T_c$. The normal scattering amplitudes are calculated from the Landau parameters. These Landau parameters are obtained from the local version of the induced interaction model for computing Landau parameters. We also calculate the leading order finite temperature correction to the diffusion lifetime. A calculation of the spin diffusion coefficient is presented in the end. Upon choosing a proper value of $F_0^a$, we are able to present a good match between the theoretical result and the experimental measurement which indicates the presence of the superfluid fluctuations near $T_c$.
\end{abstract}
\maketitle

With the first experimental realization of Bose-Einstein condensation (BEC) in a Bose gas in 1995\cite{Anderson,Bradley,Davis}, there has been an enormous amount of experimental and theoretical work carried out to study ultra-cold atomic physics\cite{BDZ,GPS}. In addition to Bose gases there are as well cold Fermi gases, with an interaction strength that can be tuned by the proximity of a Feshbach resonance. At resonance the Fermi gas is said to be at unitarity. The normal to superfluid transition temperature of an ultracold Fermi gas decreases exponentially with decreasing interaction strength of the Fermi gas in the weakly attracting limit\cite{GPS}, $T_c\approx0.28T_Fe^{\pi/2k_Fa}$. At unitarity the Fermi gas is strongly correlated, one expects a big boost in $T_c$ and thus a significantly large critical region above $T_c$ compared to a non-unitary dilute Fermi gas. The superfluid lambda transition which was once difficult to observe in a dilute Fermi gas has also been experimentally realized in a unitary Fermi gas recently\cite{Ku}.

In this paper, we are interested in revealing superfluid fluctuations in the spin diffusion lifetime of a unitary Fermi gas above $T_c$. The quasiparticle scattering amplitudes near the Fermi surface are essential in calculating the spin diffusion lifetime\cite{Pethick}. At temperatures close to $T_c$, the scattering amplitudes are greatly affected by the formation of cooper pairs. Superfluid fluctuations dominate the quasiparticle scattering right above $T_c$. The spin diffusion lifetime is calculated using Fermi-liquid theory by evaluating the total scattering probability with the superfluid fluctuations included. The Landau parameters needed in calculating the scattering amplitudes are computed using the local induced interaction model\cite{Engelbrecht,Jackiewicz}. Further, the leading order finite temperature correction to the spin diffusion lifetime is calculated to complete the calculation. Finally, the spin diffusion coefficient of a two component unitary Fermi gas is calculated to compare with the experiment\cite{Sommer}.


Superfluid fluctuations in the transport lifetimes of a unitary Fermi gas are investigated through calculating the quasiparticle scattering amplitudes of the gas near $T_c$ in a similar fashion as an earlier study on zero-sound atenuation in Liquid $^3\text{He}$\cite{Samalam}. As the temperature approaches $T_c$, the virtual formation of Cooper pairs start to dominate the quasiparticle scattering process. Singularities in the scattering amplitudes are found for small total momentum quasiparticle scattering leading to diverging scattering amplitudes at $T_c$ for zero total momentum quasiparticle scattering. The exact calculation of superfluid fluctuations in the scattering amplitudes is done by evaluating the temperature vertex function of particle-particle type in the singlet channel. The equation of the temperature vertex function is given by summing over the various ``ladder diagrams'' of the vertex function\cite{Abrikosov},
\begin{eqnarray}
\mathcal{T}_s(p_1,p_2;p_3,p_4)&=& \tilde{\mathcal{T}}_s(p_1,p_2;p_3,p_4)-\frac{T}{2(2\pi)^3} \nonumber \\ && \times\Sigma_{\omega_n}\int\tilde{\mathcal{T}}_s(p_1,p_2;k,q-k)
\mathcal{G}(q-k) \nonumber \\ && \times \: \mathcal{G}(k)\mathcal{T}_s(k,q-k;p_3,p_4)\mathrm{d}^3\mathbf{k}\label{eq:1}
\end{eqnarray}
where $\tilde{\mathcal{T}}_s$ is the temperature particle-particle irreducible vertex function, $\mathcal{G}$ is the exact temperature Green's function, $\omega_n=(2n+1)\pi T$ are the "odd frequencies" for fermions. Here we have introduced the four vector $p_i=(\mathbf{p}_i,\omega_i)$ to denote the momentums of the incident and scattered particles, and $q=(\mathbf{q},\omega_0)$ stands for the total momentum of the incident particles. $\mathcal{T}_s$ depends only on the total momentum $q$,  $\mathcal{T}_s(p_1,p_2;p_3,p_4)\equiv\mathcal{T}_s(q)$, when $|\mathbf{p}_i|=k_F$ for $i=1,\cdots,4$ and $|\mathbf{q}|\ll k_F$. Integrate out the second term on the right side of (Eq.\ref{eq:1}), we have in the small $q$ limit with $\omega_0=0$ the temperature vertex function,
\begin{equation}
\mathcal{T}_s(\mathbf{q},0)=\frac{1}{\frac{mp_f}{4\pi^2} \left[\ln\frac{T_c}{T}-\frac{1}{6}\left(\frac{v_f|\mathbf{q}|}{2\omega_D}\right)^2-\frac{7\zeta(3)}{3\pi^2} \left(\frac{v_f|\mathbf{q}|}{4T}\right)^2\right]}\label{eq:2}
\end{equation}
where $T_c=\frac{2\gamma\omega_D}{\pi}e^{-4\pi^2/mp_f|\tilde{\Gamma}_s|}$, $\mathbf{ln}\gamma$ is the Euler's constant, and $\omega_D=0.244\varepsilon_F$ is the cutoff frequency\cite{Gorkov}. Here we have set $\tilde{\mathcal{T}}_s=\tilde{\Gamma}_s$, where $\tilde{\Gamma}_s$ is the zero temperature irreducible particle-particle vertex function. $\tilde{\Gamma}_s$ is equivalent to the normal Fermi liquid scattering amplitude, $\tilde{\Gamma}_s\cdot N(0)=A_0^{Sing}=A_0^s-3A_0^a$, where $N(0)=\frac{m^*p_f}{\pi^2\hbar^3}$ is the density of states at the Fermi surface, $A_0^{s,a}=\frac{F_0^{s,a}}{1+F_0^{s,a}}$, and $F_0^{s,a}$ are the Landau parameters. As is shown by (Eq.\ref{eq:2}), the scattering amplitude at $T_c$ for zero $\mathbf{q}$ is indeed divergent. The total quasiparticle scattering probability is obtained by averaging the scattering amplitudes of different $\mathbf{q}$'s over the phase space\cite{Pethick},
$\langle W\rangle\equiv\int\frac{\mathrm{d}\Omega}{4\pi}\frac{W(\theta,\phi)}{\cos(\theta/2)}$. Superfluid fluctuations are phase space limited as virtual Cooper pair formation breaks down when the total momentum of the pair exceeds a certain value $\mathbf{q}_{max}$, where $v_f |\mathbf{q}_{max}|=\sqrt{6\varpi}$ and $\varpi=2\omega_D e^{-4\pi^2/mp_f |\tilde{\Gamma}_s|}$ from regular quantum field theory computations\cite{Abrikosov}. Hence quasiparticle scattering processes with total momentum larger than $\mathbf{q}_{max}$ are treated by normal Fermi liquid theory with the scattering amplitudes being the normal Fermi liquid scattering amplitudes. The phase space average of the scattering amplitudes could then be readily separated into a normal part and a superfluid fluctuation part,
\begin{eqnarray}
\langle W\rangle & = & \int_0^{\mathbf{q}_{max}}\frac{\mathrm{d}\Omega}{4\pi}\frac{W_f(\theta,\phi)}{\cos(\theta/2)}
+\int_{\mathbf{q}_{max}}^{2P_f}\frac{\mathrm{d}\Omega}{4\pi}\frac{W_n(\theta,\phi)}{\cos(\theta/2)}\nonumber \\
& = & \langle W\rangle_{fluctuations}+\langle W\rangle_{normal}\label{eq:3}
\end{eqnarray}
where $\langle W\rangle_{fluctuations}$ and $\langle W\rangle_{normal}$ stand for superfluid fluctuations and normal Fermi liquid scattering amplitudes respectively.

\begin{figure}[t]
 \centering
 \includegraphics[scale=0.5]{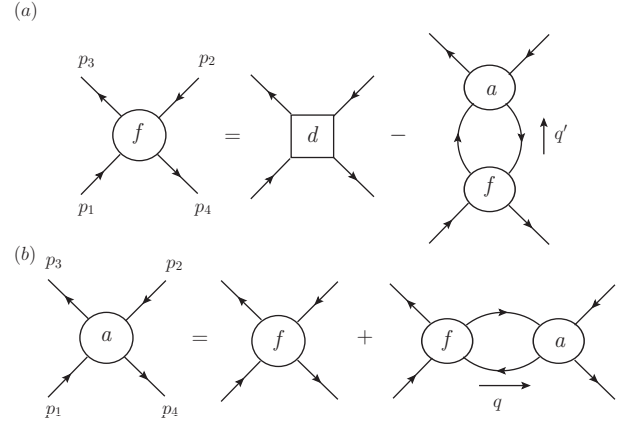}
 \caption{Diagrammatic representation of the integral equations for the Landau parameters $F$ and the scattering amplitudes $a$. (a) represents the equation for Landau parameters decomposed into direct and induced terms; (b) sums all the reducible diagrams. It represents the equation relating $F$ to the scattering amplitudes $a$.}\label{fig:diagram}
\end{figure}

The Landau parameters are calculated using local induced interaction model. The induced interaction model was first introduced in the 1970s\cite{Babu} to describe the quasiparticle interaction of liquid $^3\text{He}$. The more general momentum dependent scattering amplitude model was developed in the 1980's\cite{ABBQ,BQ,Ainsworth}. Such a theory splits the quasiparticle interaction into two pieces: the direct and the induced as shown diagrammatically in Fig. \ref{fig:diagram}. The induced term comes from the part of the interactions induced through the exchange of the collective excitations, whereas the direct term is the Fourier transform of a model dependent effective quasiparticle potential. The generalized expressions of the Landau parameters were derived diagrammatically by Ainsworth and Bedell\cite{Ainsworth}. In the local limit of a Fermi liquid the quasiparticle interaction is independent of the momentum\cite{Engelbrecht}, thus Landau parameters $F_l^{s(a)}$ with $l$ larger than $0$ are all zero. The set of equations is reduced\cite{Jackiewicz,Gaudio} to\
\begin{equation}
F_0^s = D_0^s+\frac{1}{2}F_0^sA_0^s+\frac{3}{2}F_0^aA_0^a\label{eq:6}
\end{equation}
\begin{equation}
F_0^a = D_0^a+\frac{1}{2}F_0^sA_0^s-\frac{1}{2}F_0^aA_0^a\label{eq:7}.
\end{equation}
Additionally, the scattering amplitudes are related to the Landau parameters as\cite{Pethick}
\begin{equation}
A_l^{s,a}=\frac{F_l^{s,a}}{1+F_l^{s,a}/(2l+1)}\label{eq:8}
\end{equation}
, and the forward scattering sum rule\cite{Pethick} $\sum_l(A_l^s+A_l^a)=0$ is reduced to $A_0^s+A_0^a=0$. The direct terms are fully anti-symmetrized so that $D_0^{\uparrow\uparrow(\downarrow\downarrow)}=0$. According to Ainsworth and Bedell\cite{Ainsworth}, $D_0^s=\frac{N(0)}{2}(D_0^{\uparrow\uparrow}+D_0^{\uparrow\downarrow})=\frac{2}{\pi} k_Fa_s$ and $D_0^a=-\frac{N(0)}{2}(D_0^{\uparrow\uparrow}+D_0^{\uparrow\downarrow})=-\frac{2}{\pi} k_Fa_s$, where $a_s$ is the quasiparticle s-wave scattering length. We derive from the local induced interaction model the expression for the scattering length as a function of $F_0^a$,
\begin{equation}
\frac{-1}{k_Fa_s}=\frac{8}{\pi}\frac{(1+F_0^a)(1+2F_0^a)}{F_0^a+3F_0^a(1+2F_0^a)^2}\label{eq:9}
\end{equation}
\begin{figure}
 \centering
 \includegraphics[scale=0.35]{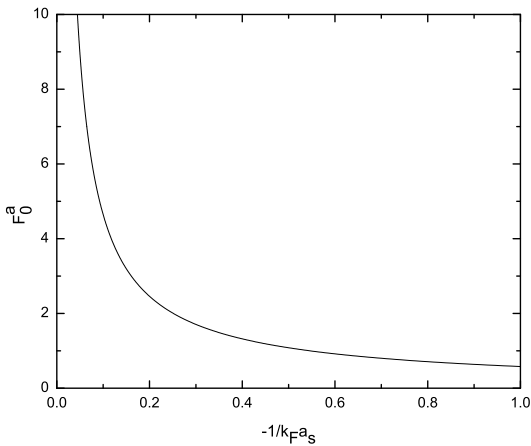}
 \caption{The Landau parameter $F_0^a$ versus $-(k_Fa_s)^{-1}$ curve.}\label{fig:graph2}
\end{figure}
This relation is depicted by Fig. \ref{fig:graph2}. The quasiparticle interaction strength of a Fermi gas is characterized by the s-wave scattering length $a_s$. On the BCS side of the BCS-BEC crossover\cite{Smith}, the s-wave scattering length of the Fermi gas is always negative and it goes to negative infinity at unitarity. Therefore, $F_0^a$ of a unitary Fermi gas approaches positive infinity. Utilizing the local induced interaction model, we are able to calculate the $F_0^{s,a}$ given the quasiparticle interaction strength of the Fermi gas.

Despite its simple structure and easy mathematics, the local induced interaction model does a good job in explaining the universal thermodynamics of a unitary Fermi gas. In a Galilean invariant system the mass renormalization disappears for a local Fermi liquid, $m^*=m$. The Landau parameter $F_0^s$ saturates to $-0.5$ at unitarity from the local model. Hence the quasiparticle mass and Landau parameter $F_0^s$ are both independent of the particle density $n$. We derive the relation between chemical potential $\mu$ and the Fermi energy $E_F$ from the regular Fermi liquid compressibility $\kappa=\frac{1}{n^2}\frac{N(0)}{1+F_0^s}$ at zero temperature,
\begin{equation}
\mu(n,0)=(1+F_0^s)E_F.\label{eq:10}
\end{equation}
We notice $F_0^s$ is equivalent to the universal parameter\cite{GPS,Ho} $\beta$ which relates the zero temperature chemical potential to the Fermi energy of a unitary Fermi gas. The local induced interaction model\cite{Jackiewicz,Gaudio} gives the value of $F_0^s=-0.5$ at unitarity. The value falls close to various experimental results of $\beta$ ranging from $-0.73$ to $-0.49$ and Quantum Monte Carlo result\cite{GPS} $-0.58$. The leading order temperature dependence of several thermodynamic quantities is studied using basic thermodynamic analysis. We introduce the temperature scale $T_s\equiv\mu(n,0)$, in the absence of spin polarization and mass renormalization, the chemical potential of a Fermi gas is given as\cite{Pethick},
\begin{equation}
\mu(n,T)=\mu(n,0)\left[1-\frac{\pi^2(1+F_0^s)}{12}\left(\frac{T}{T_s}\right)^2\right].
\end{equation}
The chemical potential scales the same in temperature as a free Fermi gas. The total entropy is given as $S/N k_B=\frac{\pi^2(1+F_0^s)}{2}\frac{T}{T_s}$ from Fermi liquid theory. Based on thermodynamic relations $\kappa=\frac{1}{n^2}\frac{\partial n}{\partial \mu}$ and $dP=nd\mu+sdT$, we calculate the compressibility to be $\kappa(n,T)=\kappa(n,0)\left[1+\frac{\pi^2(1+F_0^s)}{12}\left(\frac{T}{T_s}\right)^2\right]^{-1}$ and the pressure to be $P(n,T)=P(n,0)\left[1+\frac{5 \pi^2 (1+F_0^s)}{12}\left(\frac{T}{T_s}\right)^2\right]$ where $\kappa(n,0)=\frac{1}{n^2}\frac{N(0)}{1+F_0^s}$ and $P(n,0)=\frac{2}{5}(1+F_0^s)nE_F$. All the thermodynamic quantities calculated above involve universal functions of the Fermi energy $E_F$ and the ratio $T/T_s$ as expected for a unitary Fermi gas\cite{Ho}. In addition, an effective s-wave scattering amplitude $\tilde{a}_0$ could be defined as,\cite{Abrikosov}
\begin{equation}
\frac{\tilde{\Gamma}_s}{2}=\frac{4\pi \tilde{a}_0}{m}.\label{eq:11}
\end{equation}
Analogous to the two body scattering problem, we can write down the s-wave phase shift\cite{Landau}
\begin{equation}
\delta_0=k_F \tilde{a}_0=\frac{\pi}{8}A_0^{Sing}. \label{eq:12}
\end{equation}
Using the local induced interaction model, we show $\delta_0=-\frac{\pi}{2}$ on the BCS side of BCS-BEC crossover and $\delta_0=\frac{\pi}{2}$ on the BEC side of the crossover. Based on Levinson's theorem, the increase in the phase shift by $\pi$ indicates the appearance of a bound state on the BEC side of the BCS-BEC crossover in agreement with the physics of the BCS-BEC crossover. A rough estimate of the molecular binding energy on the BEC side is made using $E_b=-\hbar^2/m\tilde{a}_0^2\approx-0.8E_F$. We can also calculate the superfluid transition temperature of a unitary Fermi gas. According to the local induced interaction model, $F_0^a\to\infty$, hence $A_0^a=1$ in the unitary limit. The superfluid transition temperature is then given\cite{Gorkov} by $T_c=0.28T_Fe^{-4\pi^2/mp_f|\tilde{\Gamma}_s|}=0.28T_Fe^{-1/|A_0^a|}=0.102T_F$. The local model predicts a $T_c$ value relatively close to the experimentally measured\cite{Ku} $T_c=0.167T_F$. In the later calculations, we introduce a scaling factor $L=1.64$ in the exponential term of $T_c$, $e^{-1/|A_0^a|}\to Le^{-1/|A_0^a|}$, to artificially lift $T_c$ from the local model to its experimental value.

The calculation of the spin diffusion lifetime $\tau_D$ is straight forward. In the low temperature limit $T\ll T_F$, it can be formulated in the language of Landau Fermi-liquid theory\cite{Pethick}. $\tau_D$ is proportional to the characteristic relaxation time $\tau$ defined as, \begin{equation}
\tau\equiv\frac{8\pi^4\hbar^6}{m^{\ast3}\langle W\rangle (k_BT)^2}
\end{equation}
where $\langle W\rangle\equiv\int\frac{\mathrm{d}\Omega}{4\pi}\frac{W(\theta,\phi)}{\cos(\theta/2)}$ and $W(\theta,\phi)=\frac{1}{2}(\frac{1}{2}W_{\uparrow\uparrow}+W_{\uparrow\downarrow})=\frac{1}{2}W_{\uparrow\downarrow}(\theta,\phi)$ is the average scattering probability. The triplet scattering amplitude is zero in the local limit. Taking into consideration of the superfluid fluctuations, the scattering probabilities $W_f(\theta,\phi)$ and $W_n(\theta,\phi)$ are calculated from their respective scattering amplitudes,
\begin{equation}
W_f(\theta,\phi)=\frac{1}{2}W_{\uparrow\downarrow}= \frac{1}{2}\frac{2\pi}{\hbar}|t_{\uparrow\downarrow}|^2 = \frac{1}{2}\frac{2\pi}{\hbar}\Big|\frac{\mathcal{T}_s(\mathbf{q},0)}{2}\Big|^2 \label{eq:14}
\end{equation}
\begin{equation}
W_n(\theta,\phi)=\frac{1}{2}W_{\uparrow\downarrow}=\frac{1}{2}\frac{2\pi}{\hbar}|t_{\uparrow\downarrow}|^2
=\frac{1}{2}\frac{2\pi}{\hbar}\Big|\frac{-2A_0^a}{N(0)}\Big|^2\label{eq:15}
\end{equation}
Performing the integrals in (Eq.\ref{eq:3}), we have the phase space average of the scattering amplitudes,
\begin{equation}
\langle W \rangle_{normal}= \frac{2\pi}{\hbar}\frac{2}{|N(0)|^2}\cdot 2(1-\frac{\sqrt{6}\pi}{4\gamma}\frac{T_c}{T_F})|A_0^a|^2 \label{eq:16}
\end{equation}
\begin{widetext}
\begin{eqnarray}
&&\langle W \rangle_{fluctuation} = \frac{2\pi}{\hbar}\frac{2}{|N(0)|^2}  \nonumber \\ && \times \left[ \frac{\frac{\sqrt{6}\pi T_c}{4\gamma T_F}}{\ln\frac{T}{T_c} \left[ \ln\frac{T}{T_c}+ \left(\frac{\sqrt{6}\pi T_c}{4\gamma T_F}\right)^2 \left(11.2+0.28 \left(\frac{T_F}{T_c}\right)^2 \right)\right]} +\frac{\tan^{-1}\left(\sqrt{\left(\frac{\sqrt{6}\pi T_c}{4\gamma T_F}\right)^2\left(11.2+0.28 \left(\frac{T_F}{T_c}\right)^2 \right)}/\sqrt{\ln\frac{T}{T_c}}\right)}{\left(\ln\frac{T}{T_c}\right)^{3/2} \sqrt{11.2+0.28 \left(\frac{T_F}{T_c}\right)^2}}\right] \label{eq:17}
\end{eqnarray}
\end{widetext}
The low temperature expression of the spin diffusion lifetime is readily given by,
\begin{equation}
\tau_D^0=\left(\frac{\tau_D}{\tau}\right)\tau=\frac{0.129\times8\pi^4\hbar^6}{m^3\langle W\rangle(k_BT)^2}.\label{eq:18}
\end{equation}
The leading order finite temperature correction to $\tau_D^0$ is computed\cite{Dy},
\begin{eqnarray}
\frac{1}{\tau_D} &-& \frac{1}{\tau_D^0}=-\frac{3}{2}\pi\zeta(3)\frac{k_B T_F}{\hbar}\left(\frac{T}{T_F}\right)^3\nonumber \\
&\times&[-2.95(A_0^a)^3+1.564(A_0^a)^2+1.278A_0^aF_0^a]\label{eq:19}
\end{eqnarray}
We obtain the full expression of $\tau_D$ by solving (Eq.\ref{eq:19}),
\begin{eqnarray}
\tau_D &=& \frac{\hbar}{k_B T_F}\left(\frac{T_F}{T}\right)^2\bigg(\frac{\hbar|N(0)|^2}{0.129\times16}\langle W\rangle-\frac{3}{2}\pi\zeta(3)\nonumber \\ &\times&[-2.95(A_0^a)^3+1.564(A_0^a)^2+1.278A_0^aF_0^a]\frac{T}{T_F}\bigg)^{-1}\label{eq:20}
\end{eqnarray}

In the end, we calculate the spin diffusion coefficient of a Fermi gas from $\tau_D$ to compare with the experiment\cite{Sommer}. An estimate of  the high temperature limit ($T\gg T_F$) of $\tau_D$ is included. The transport lifetime scale as $\tau\propto\frac{\hbar}{k_B T_F}\left(\frac{T}{T_F}\right)^{1/2}$ at high temperature\cite{Bruun}. We choose the numerical factor in front to be $5.84$ such that $\tau_D\approx 5.84\times\frac{\hbar}{k_B T_F}\left(\frac{T}{T_F}\right)^{1/2}$ is consistent with the high temperature portion of experimentally measured spin diffusion coefficient\cite{Sommer}. The spin diffusion coefficient is expressed as,
\begin{equation}
 D = \left\{
  \begin{array}{ll}
   \frac{1}{3} v_f^2(1+F_0^a)\tau_D & \text{for } T\ll T_F, \\
   \\
   \frac{k_B T}{m}\tau_D & \text{for } T\gg T_F.\label{eq:25}
  \end{array} \right.
\end{equation}

\begin{figure}
 \centering
 \includegraphics[scale=0.55]{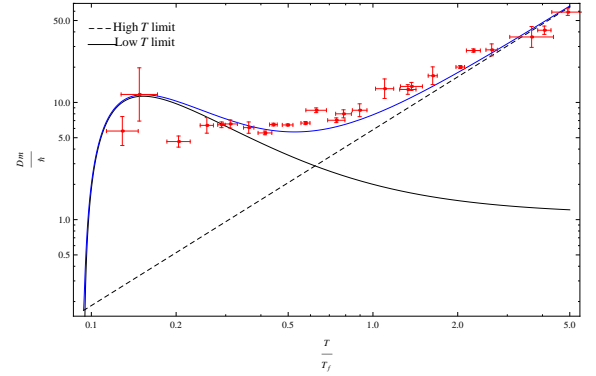}
 \caption{Spin diffusion coefficient curve from low to high temperature. The black solid curve is the low temperature expansion of $D$ at $F_0^a=1.7$; the dashed line is the classical limit of $D$; the blue curve represents the summation of the two limits; the red dots with error bars are the experimental data.}\label{fig:Diffusivity}
\end{figure}

\begin{figure}[t]
 \centering
 \includegraphics[scale=0.4]{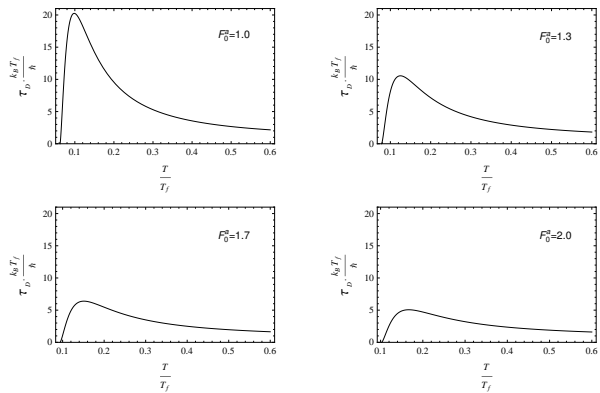}
 \caption{The calculated spin diffusion lifetime curves with different $F_0^a$ values.}\label{fig:DGraph}
\end{figure}

There exists a singularity in $\tau_D$ when the temperature increases to a point $T^*$ where the finite temperature correction term becomes comparable to $1/\tau_D^0$ according to (Eq.\ref{eq:20}). This singularity is an artifact of overextending the correction term in temperature and can be removed by expanding $\tau_D$ up to second order in $T_F/T$ in (Eq.\ref{eq:20}). The low temperature feature of $\tau_D$ is well approximated by the expansion for $T\ll T^*$. We use the expansion to describe the low temperature behavior of $\tau_D$ hereafter. By choosing $F_0^a=1.7$ and plotting the sum of the classical and low temperature limit of the spin diffusion coefficient given by (Eq.\ref{eq:25}), we are able to present a good match between the calculated and the measured spin diffusion coefficient\cite{Sommer} as depicted by Fig. \ref{fig:Diffusivity}. The use of the low temperature expansion of $\tau_D$ is justified since the low temperature portion of the experimental data we are trying to describe falls below $T^*$ ($T^*\approx0.65T_F$ for $F_0^a=1.7$). The superfluid fluctuations cause the spin diffusion coefficient $D$ to drop drastically when temperature approaches $T_c$. As the temperature moves away from $T_c$, $D$ exhibits normal Fermi-liquid like behavior going as $1/T^2$ for $T_c<T\ll T_F$.
We plot $\tau_D$ with respect to $T/T_F$ for several different values of $F_0^a$ to see how $\tau_D$ evolves with different choices of $F_0^a$. The result is depicted by Fig. \ref{fig:DGraph}. The height of the peak in $\tau_D$ decreases as $F_0^a$ increases which indicates that the superfluid fluctuations start to dominate at a higher temperature for a bigger $F_0^a$. This is expected since $T_c$ increases when $F_0^a$ increases. The theory fails to capture the correct feature of $\tau_D$ at intermediate temperatures when $F_0^a$ becomes too large, but it succeeds in revealing the superfluid fluctuations above $T_c$ through $\tau_D$ regardless of the choice of $F_0^a$.

We have introduced the local approximation for the Fermi liquid description of the cold atom Fermi gases and used the local version of the induced interaction to calculate the Fermi liquid parameters.  This has been done since this provides simple analytic results that provide qualitative and reasonably good quantitative results for the Fermi liquid parameter, $F_0^s$, and the thermodynamic scaling temperature, $T_s$, as well as $T_c$.  In earlier publications\cite{Gaudio,Gaudio2007} we used the momentum dependent induced interaction which generated Fermi liquid parameters with $l > 0$.  In the unitary limit the induced interaction gives a small mass correction, about $15\%$ above the bare mass, and it gives an $F_0^s = -0.6$.  These numbers are independent of the density at unitarity so the thermodynamic scaling is just like what we found for the local model but with a smaller value for $T_s$.  We also found that when we use the s-p approximation\cite{Pethick,Gaudio2007} to construct the scattering amplitude from our Fermi liquid parameters we get better fits for some of our calculated properties. These include, $T_c$ and $E_b$, where $T_c\approx0.14T_F$ and $E_b\approx-0.3E_F$. Clearly, we can get better numerical results going beyond the local model but it would not give us qualitatively new insights into some of the properties of this cold atom system.  In particular this would not qualitatively change the nature of the strong superfluid fluctuation effects in the spin diffusion just above $T_c$.

In conclusion, we have presented a complete formula of calculating the transport lifetime above $T_c$ of a Fermi gas with arbitrary quasiparticle interaction strength through control of $F_0^a$. Superfluid fluctuations above $T_c$ in a unitary Fermi gas are revealed through calculation of the spin diffusion lifetime. Sudden decreases in $\tau_D$ above $T_c$ are found as the evidence of the superfluid fluctuations. Upon choosing a proper value of $F_0^a=1.7$, we are able to describe the experimental data of the spin diffusion coefficient using our theory. Similar analysis will be performed to the viscous lifetime and thermal diffusion lifetime in a later paper. Further work could also be done by using the s-p approximation with the induced interaction model for calculating the Landau parameters.

Acknowledgments: We thank, J. Engelbrecht, S. Gaudio, B. Mihaila, P. Souders, J. Thomas, E. Timmermans, and M. Zwierlein, for valuable and insightful discussions.

\end{document}